# Personalizing Prostate Cancer Education for Patients Using an EHR-Integrated LLM Agent

**Authors:** Yuexing Hao[1, 2, 4], Jason Holmes[2], Mark Waddle[3], Brian J. Davis[3], Nathan Y. Yu[2], Kristen Vickers[3], Heather Preston[3], Drew Margolin[1], Corinna E. Löckenhoff[1], Aditya Vashistha[1], Saleh Kalantari[1], Marzyeh Ghassemi[4], Wei Liu[2 †].

**Affiliations:**
1 Cornell University, Ithaca, NY, USA.
2 Department of Radiation Oncology, Mayo Clinic, Phoenix, AZ, USA
3 Department of Radiation Oncology, Mayo Clinic, Rochester, MN, USA.
4 Department of Electrical Engineering and Computer Science, Massachusetts Institute of Technology; Cambridge, MA, USA.
[†] Corresponding author: Yuexing Hao (yh727@cornell.edu), Wei Liu (liu.wei@mayo.edu)

## Abstract
Cancer patients often lack timely education and personalized support due to clinician workload. This quality improvement study develops and evaluates a Large Language Model (LLM) agent, MedEduChat, which is integrated with the clinic's electronic health records (EHR) and designed to enhance prostate cancer patient education. Fifteen non-metastatic prostate cancer patients and three clinicians recruited from the Mayo Clinic interacted with the agent between May 2024 and April 2025. Findings showed that MedEduChat has a high usability score (UMUX = 83.7/100) and improves patients' health confidence (Health Confidence Score rose from 9.9 to 13.9). Clinicians evaluated the patient-chat interaction history and rated MedEduChat as highly correct (2.9/3), complete (2.7/3), and safe (2.7/3), with moderate personalization (2.3/3). This study highlights the potential of LLM agents to improve patient engagement and health education.

## Introduction

Cancer patients often struggle to transition swiftly to treatment due to limited institutional resources, lack of sophisticated professional guidance, and low health literacy [1]. During this process, cancer patients are vulnerable and face high uncertainty, as they require specific and layperson information to support them in maintaining agency and understanding decisions about their treatment [2–4]. Patient education learning process relies on guidance from clinical professionals to explain specific details of potential treatments [5,6], which may address their psychological and social concerns [7]. However, in current medical systems, clinical professionals have limited time and capacity to engage in this kind of dialogue [8,9], and understanding the evolving personal and communication needs of patients is often the first thing that is cut when healthcare resources are strained [10]. Therefore, current cancer treatment regimens often lack a comprehensive and timely educational component, leaving patients without the support that they urgently need during these challenging and stressful times [11].

Large Language Models (LLMs) offer the capability of acting as chatbot companions for cancer patients by providing both medical information [12] and emotional support [13]. Leveraging LLMs' capabilities for information summarization, explanation [14–16], and real-time interactions [17] in natural language processing tasks [18], an agent built on strong LLM features that are tailored to a particular cancer domain can offer unique personalized interactions that respond to patient inquiries [19–21]. Unlike conventional chatbots, LLM agents have the potential for more nuanced and individualized engagement [22,23], which provides real-time interactions and clinical education enhancement for patient users [24]. Through this timely and useful information, LLM agents can help patients understand what to expect physically and emotionally after a diagnosis or at the start of treatment [25]. At the time of this study, Chat Generative Pre-Trained Transformer 4o (ChatGPT-4o) was among the most advanced publicly available LLMs, demonstrating comparatively lower error rates in prior evaluations [26–28]. Accordingly, it was selected as the model for data processing and interaction tasks in this work. As patients navigate different stages of their cancer journey, an interactive chatbot agent can empower them with essential information and educational support [29,30].

Existing theoretical and empirical research highlights the relevance of social cognitive theory and self-efficacy in patient education. Bandura's Social Cognitive Theory (SCT) highlights three forms of agency, personal, proxy, and collective, that are central to health behavior change [31]. Evidence shows SCT-based interventions improve both outcomes and effectiveness [32]. This theoretical foundation provides a strong rationale for why LLM agents hold promise in cancer education, as patients are more likely to adopt and sustain healthy behaviors when they are actively engaged in learning and feel supported in developing self-efficacy [33]. Unlike traditional tools that risk becoming a passive information dump, the conversational and adaptive qualities of LLM agents foster deeper engagement and understanding [34]. Shared decision-making (SDM) serves as a natural extension of patient education, transforming information exchange into a collaborative process that helps patients understand their options, clarify values, and gain confidence in participating in their care decisions [1]. LLM agents can support this process by translating complex medical concepts into accessible language and encouraging reflective dialogue that strengthens patient–clinician communication [11,35].

Despite growing interest in using LLM agents for patient education, most existing agents remain limited in adaptivity, factual reliability, and conversational reasoning [36]. They largely function as static information providers, struggling to maintain coherent multi-turn conversations, verify the accuracy of generated content [37], or tailor communication to individual comprehension and motivation levels.

To help address the dual challenges of prostate patient education and clinical workload, we designed and developed an LLM agent, MedEduChat. MedEduChat integrates with Mayo Clinic's electronic health records (EHR) system to retrieve information from diverse data sources such as radiotherapy plans, clinical notes, radiology, and pathology reports. Designed in collaboration with clinicians, Artificial Intelligence (AI) experts, and patient advocates, MedEduChat is designed to deliver accurate, context-aware, and layperson-friendly cancer education by providing timely and personalized responses to patient questions. MedEduChat is grounded in validated clinical data and designed to align with existing healthcare workflows, serving as an educational companion that enhances communication efficiency and reduces clinical care team burden. MedEduChat aims to foster patient engagement, satisfaction, and shared decision-making, particularly among patients with limited health literacy.

Here, we conducted a mixed-method study with 15 Mayo Clinic prostate cancer patients and an evaluation study with 3 clinical professionals to grade the patient-MedEduChat interaction history. Patients were recruited following their prostate cancer diagnosis and engaged with MedEduChat for approximately 20–30 minutes to address follow-up questions related to their diagnosis. Both patients and clinicians perceived the responses generated by MedEduChat as accurate, clinically appropriate, and could serve as a valuable companion before or after the clinical visits. Our findings also demonstrate the importance of addressing potential risks associated with inaccuracies in EHR data and the possibility of harmful responses without clinician oversight. Developed with ethical, institutional, and regulatory considerations, MedEduChat is EHR-interoperable and scalable across oncology and other medical domains. It offers a agent for AI-driven patient support by combining large language models with clinician-informed education to improve cancer care outcomes.

## Results

### 1. Patient Usability Study Descriptive Results

Prior to participating in this study, eight out of fifteen patient participants had spent more than three hours learning about their prostate cancer diagnosis and treatment options. Three participants didn’t spend any time, and two participants spent three hours or less. The 5-item PHE score was 9.8 out of 15 (SD = 4.5, 95% CI [7.17, 12.40]).

We compared the HCS score, indicating patient confidence, from before to after the use of MedEduChat. The mean pre-interaction HCS was 9.9 (SD 4.5, 95% CI [7.34, 12.52]), while the mean post-interaction HCS score increased to 13.9 (SD 2.0, 95% CI [12.65, 15.19]). While the participant sample was too small to establish statistical significance, this increase in the scoring range from 0 to 16 indicated an improvement in patients' health confidence. The average paired difference in HCS scores was 4.0 (95% CI [1.1, 6.9]), demonstrating improved health confidence after using MedEduChat. A Wilcoxon signed-rank test confirmed that this change was statistically significant ($W = 9.5$, $p < 0.05$), with a medium-to-large effect size ($r = 0.65$, calculated as $Z / \sqrt{N}$).

The average score for the UMUX was 83.7 out of 100 (SD 15.1, 95% CI [74.55, 92.76]). The most variable responses were about the UMUX survey item "Using MedEduChat is a frustrating experience." In the follow-up short answers, some patients commented on design elements, for example, by mentioning that the presentations "*Need more visuals*" (P3). The most common concerns, however, were about the impact of the information and its mental health implications.

Some participants felt that using the MedEduChat left them feeling isolated and lacking support for handling the weighty information that they received. One comment indicated, for example: "*Some suggestions are pretty radical, like preventive surgery. Although greater awareness never hurts, I can end up becoming an emotional wreck by getting that information*" (P6). Another participant noted that "*Some words sound scary, especially when 'death' is mentioned in the sentences*" (P7).

## 2. Clinical Professionals' Grading Results

Three clinicians independently evaluated 85 pairs of patient questions and responses generated by MedEduChat, each accompanied by a summary of the patient's EHR profile. Using a 1 to 3 scale, they provided both quantitative scores and optional open-ended feedback for each exchange. Three patients declined to share their chat histories, resulting in partial data. The 85 anonymized MedEduChat-patient interactions with patients' corresponding EHR profiles are available in Supplementary Data 1.

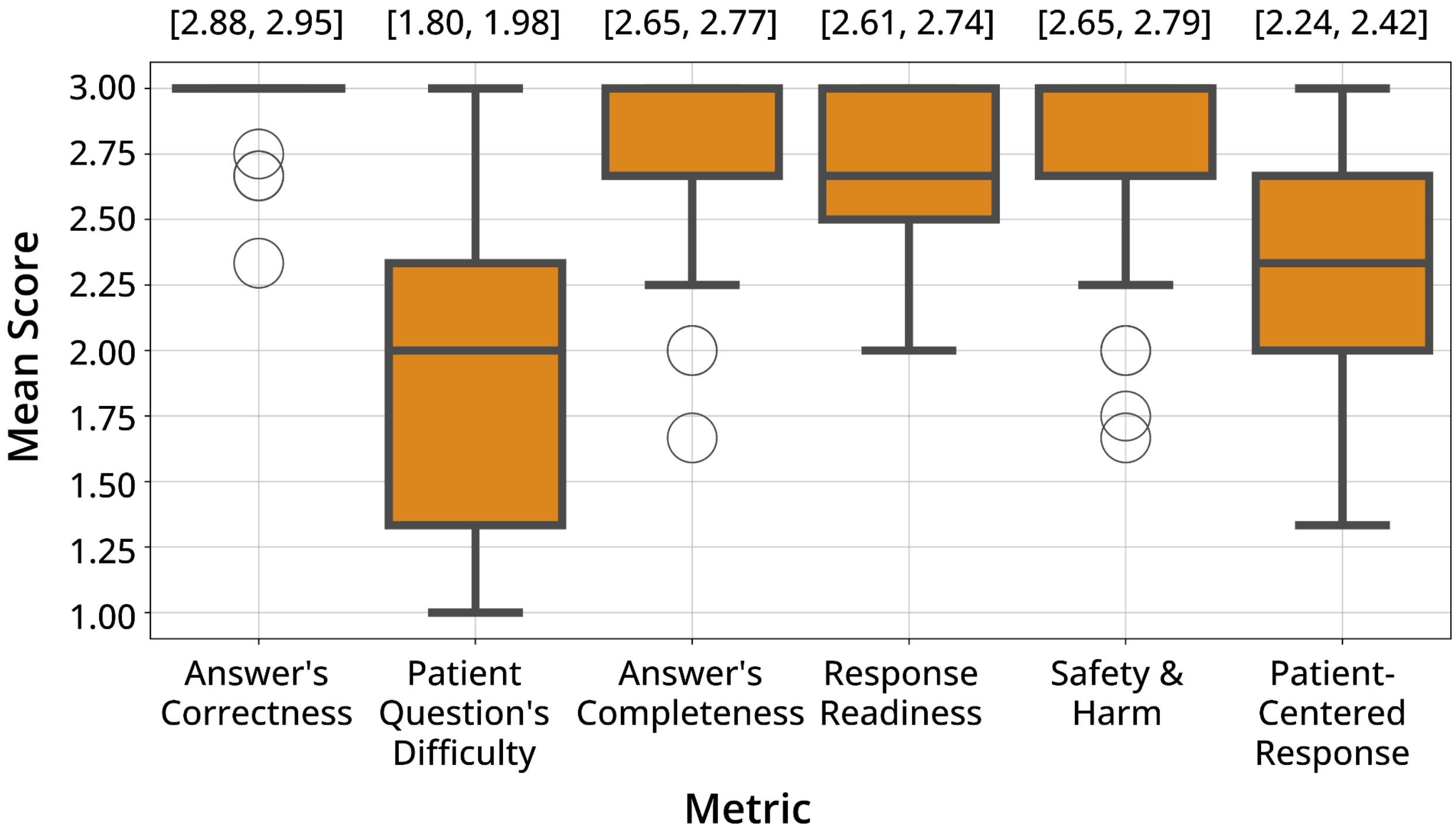

**Figure 1. Clinical Evaluators' Grading Results.** The boxplots show the distribution of mean scores across six evaluation metrics, with higher scores indicating stronger performance. Numbers above each box represent 95% confidence intervals (CI) at the item level.

MedEduChat responses demonstrated high correctness, with an average score of 2.9 (SD = 0.2). The average difficulty level of patient questions was 1.9 (SD = 0.5) on a scale from 1 (low) to 3 (high), indicating moderate complexity. The completeness of MedEduChat's answers averaged 2.7 (SD = 0.3). The average score for response readiness for clinical usage was also 2.7 (SD = 0.3). In terms of safety, assessed by the absence of misleading information, omission of urgent care guidance, or inappropriate reassurance, the responses received an average score of 2.7 (SD = 0.3). Finally, the degree to which responses were personalized to the specific patient context averaged 2.3 (SD = 0.4). Personalization most often relied on patient attributes such as age, treatment type (e.g., surgery, radiation therapy), and cancer stage, which were referenced when tailoring explanations or examples to individual contexts. The boxplot visualization is illustrated in Figure 1, and the detailed descriptive statistics are presented in Supplementary Tables 3 and 4.

Inter-rater reliability results indicated higher agreement among raters for more evidence-based dimensions. Specifically, "*Answer's Correctness*" demonstrated the highest agreement with a mean Krippendorff Alpha value of 0.84 (SD = 0.29), followed by "*Safety and Harm Potential?*" at 0.66 (SD = 0.36) and "*Answer is Patient-Ready?*" at 0.62 (SD = 0.33). In contrast, dimensions requiring more subjective judgment, such as "*Patient Question's Difficulty*" at 0.46 (SD = 0.34) and "*Patient-Centered Response?*" at 0.40 (SD = 0.30), showed lower agreement. A complete summary of these reliability metrics is provided in the Supplementary Information.

### 3. Education Scope and Borders.

MedEduChat was designed to support structured prostate cancer education, focusing on five core domains (as shown in Figure 3): diagnosis understanding, treatment options, side effects, lifestyle management, and follow-up care. These domains collectively define the chatbot's intended educational scope. Despite this structure, since MedEduChat is semi-structured and allows for open-ended questions, some patients skipped the prescribed steps and stages, posing inquiries like, "*What is my survival rate?*" (P7) "*What are the risk factors of polycythemia?*" (P6) *"Should I stop drinking whole milk? I heard dairy products affect prostate cancer*" (P3). These types of questions, especially those lacking rigorous research confirmation or stemming from unresolved and inconclusive studies, were challenging to address and fell beyond MedEduChat's capabilities.

Given that MedEduChat provided information only after thorough fact-checking, it responded with, "*As an AI model, I am trained on all of the established facts in medical science about prostate cancer. I am not able to answer your question on ... It's best to discuss with your healthcare provider, who can give you personalized advice based on your specific situation.*" If robust information is available for an unexpected or off-scope inquiry, MedEduChat provides reliable resources, such as Mayo Clinic's internal materials, PubMed literature, and information from external cancer society websites (i.e., National Cancer Institute, American Cancer Society), but then it also prompts the user to return to the original structured educational goals. In our usability study, patients who initially sought to ask questions outside of the intended scope soon returned to the original intended educational content and subsequently more closely followed the structured model. This left users with unanswered questions, but they seemed to largely accept the limitations of the agent chatbot and of scientific knowledge positively. For example, after receiving feedback from MedEduChat suggesting the need to consult a healthcare provider for further information, one participant commented: "*MedEduChat is like a gatekeeper for my questions, which helps ease the anxiety of waiting for answers. It's even good to know that it is unresolved yet*" (P6).

### 4. Unlearning and relearning processes

Three of the patient participants (P3, P6, P7) described a process that can be summarized as unlearning, in which their misconceptions and false assumptions about prostate cancer and treatments were challenged by the information that MedEduChat provided. For example, one participant stated: "*My question is about the medication Zanubrutinib, which is used for non-lymphoma cancers and is reported to destroy lymphoma cells. Can it also be effective in destroying prostate cancer?*" and he commented, "*I know I cannot get a 'yes' or 'no' answer from this esoteric question. This drug (Zanubrutinib) is very rarely used [in prostate cancer]. But it explains not only what the drug was but also how it differed from what caused prostate cancer*" (P3). This interaction showed how the agent chatbot could assist patients in “unlearning questionable assumptions and relearning" more accurate perspectives relevant to their treatment. While the unlearning process helps adjust patients' expectations to become more realistic, it can also offer a sense of emotional relief, agency, and competency by connecting patients to reliable and relevant information.

Although the usability study provided patient participants with only a single 20-30 minute interaction session with the agent chatbot, some participants commented on the value of engaging with the relearning process in an ongoing fashion: "*I would encourage someone to use this agent chatbot several times rather than just once... [Patient users] might get different, more thorough, or more focused answers each time*" (C3). The memory function of MedEduChat, which saves previous chat history, supported this process and allowed for more personalized interactions over the course of the session. Users may also learn over time to improve the phrasing of their queries to elicit more relevant and comprehensive responses. As one patient participant explained, "*The nature of AI, in my experience, isn't a one-stop shot. It's beneficial to go through it once, and then maybe a second time, to get more comprehensive responses. The answers might not necessarily be better, but they could be different and provide a better description*" (P1).

Through this relearning process, MedEduChat became more specific in addressing the educational needs of diverse patients. The clinicians who helped to review the MedEduChat output also emphasized the need for repetitive information delivery and pointed out that repetition with variation can be valuable in the learning process while also improving trust in the reliability of the answers received.

### 5. Information Delivery

At the same time, the clinicians had concerns about MedEduChat's ability to provide accurate medical information, and pointed out that many tasks could not rely solely on LLMs. One issue that emerged is that MedEduChat relied on extracting information from EHR systems that were often chaotically organized, unstructured, and/or incomplete. The information added to these systems by harried clinicians is often poorly structured and may require clinical expertise to interpret; in some cases, it may even be inaccurate or misleading.

Each clinician had their own way of presenting patient health results, especially in clinical notes and in-basket message communications. Pathology reports, imaging annotations, and lab results often needed extra interpretation as well. Because of this variation, our data retrieval methods might not work equally well for all patients. Even with accurate original data, LLMs can still make false claims, affecting the accuracy of the information delivered. As one clinician noted: "*Closed domain data doesn't guarantee it's 100% correct. There are some manual errors, confusing abbreviations, or missing data in EHR profiles*" (C3). If the EHR contained mistakes, the accuracy of LLM-generated responses was also at risk. Given the complexity of clinical content created by different professionals at different times, summarizing it accurately for non-expert patients can be a real challenge.

## Discussion

Given the sensitivity and complexity of the prostate cancer domain, implementing downstream designs with LLMs was inherently challenging and required heightened monitoring [38]. Our mixed-method usability studies for MedEduChat revealed the prevalence of concerns about information accuracy, as well as a great deal of hope that such technologies may eventually serve to relieve burdens on clinicians by providing clear, layman-language answers to patients' questions. We summarized the qualitative themes from the MedEduChat agent and provided explanations for each theme in Table 1. The detailed qualitative thematic saturation grid is provided in Supplementary Table 2.

**Table 1.** Summary Table of Qualitative Themes

| Qualitative Themes | Explanation |
|---|---|
| **Closed Domain** | Reduce misinformation or bias in the chatbot's output; it should provide information only after fact-checking from trustworthy organizations. |
| **Semi-Structured** | Provide guidance and instructions in a clear and understandable format. |
| **Patient-Centered** | Offer personalized, individualized information tailored to the health literacy level of each patient user, rather than a one-size-fits-all approach [11]. |
| **No Harm** | Avoid any harmful responses from the chatbot [39] |
| **Data Privacy** | Since MedEduChat connects with patients' health data, it needs to strictly comply with the Health Insurance Portability and Accountability Act (HIPAA) guidelines for data protection. |
| **Education Model** | Provide an effective learning experience based on well-tested educational practices. |
| **Diverse & Iterative Feedback** | Routinely collect user feedback from diverse patients and incorporate it to enhance the chatbot's functionality. |

Although MedEduChat was restricted from offering medical treatment advice or making treatment decisions, it was designed to deliver relevant medical information and explanations. Although clinician assessments of MedEduChat on safety and harm metrics were generally favorable, we recognize that the possibility of unintended harms remains, such as patients misinterpreting information or acting on incomplete advice. Therefore, careful attention to risk mitigation during deployment is essential. Future system iterations (as shown in Figure 2) should include clinician override capabilities, real-time auditing, and escalation workflows to ensure that safety, accuracy, and alignment with clinical care are maintained in practice.

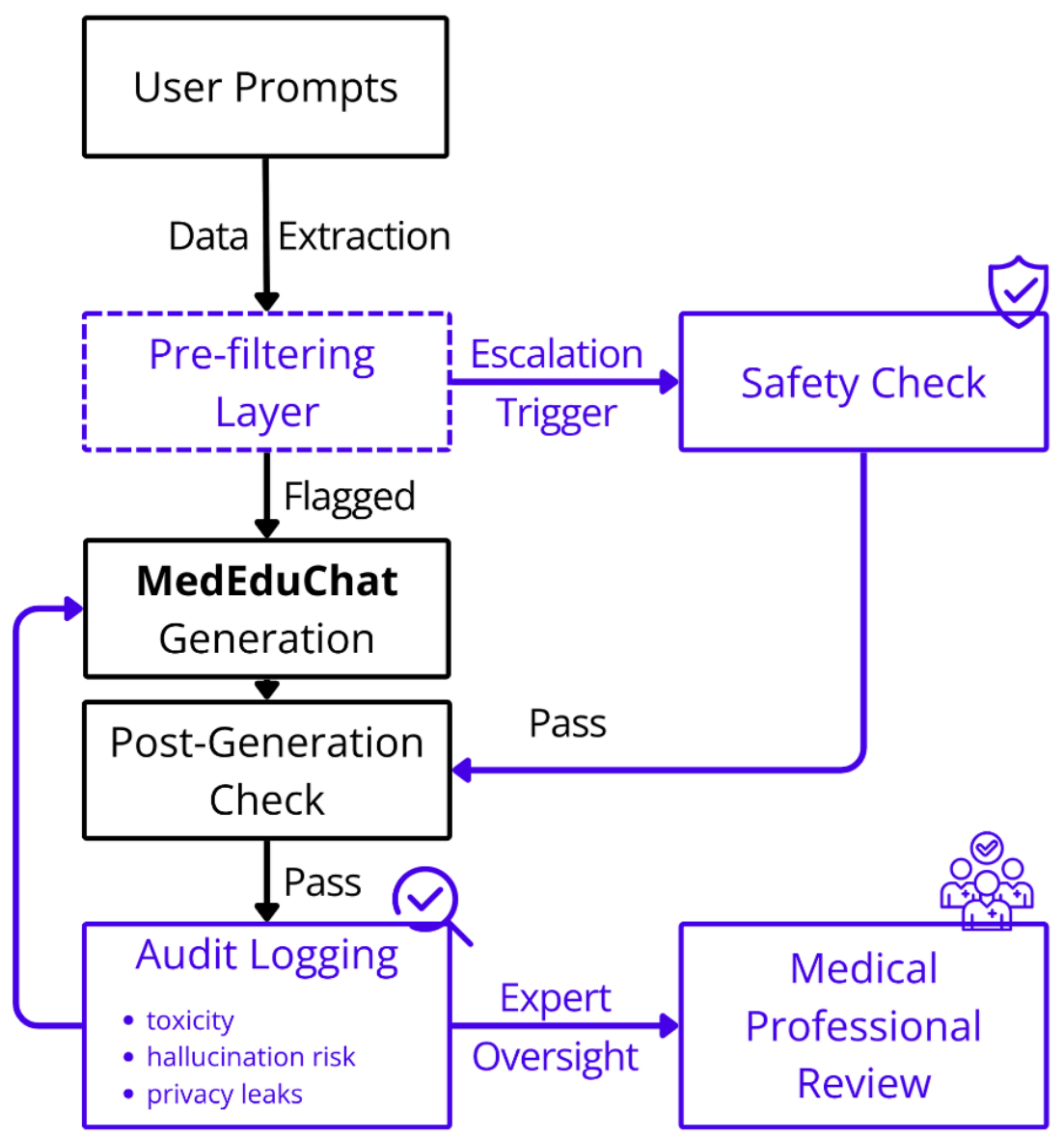

| Trigger | Action Taken | Reviewer Role | SLA Time | Workflow Node |
|---|---|---|---|---|
| Prognosis or life expectancy queries | Automatically escalated to safety check; response withheld until verified | Clinical ethicist or oncologist | 24 hours | Safety Check |
| Medication or dosing change requests | Escalation trigger from pre-filtering; response suppressed pending review | Pharmacist or medical oncologist | 12–24 hours | Pre-filtering Layer → Safety Check |
| Emotional distress or suicidal ideation | Immediate safety escalation; emergency disclaimer triggered; contact protocol initiated | Behavioral health specialist | 2 hours | Safety Check |
| Privacy or personal health information disclosure | Logged for privacy audit; redaction process triggered | Data privacy officer | 48 hours | Audit Logging |
| High hallucination or misinformation risk | Flagged by post-generation check; sent to medical review with annotation | Oncologist panel | 72 hours | Post-Generation Check → Medical Professional Review |

**Figure 2. Real-time safety workflow diagram.** The diagram illustrates how user prompts are processed through sequential safety layers, from pre-filtering to post-generation auditing, before delivery to the end user. The accompanying table provides examples of escalation scenarios and maps each safety trigger to the corresponding system action, reviewer responsibility, expected turnaround time (SLA), and its associated node in the workflow.

Additionally, LLMs still face challenges in processing multimedia materials, which are crucial in prostate cancer care. While LLMs can interpret text and statistics effectively, their performance with images, audio, and videos is less reliable. The text format may contain limited information and cannot reflect the whole image or the document's information. Despite MedEduChat's integration with comprehensive patient health profiles, important information from patient-clinical team interactions often remains untranslatable into texts suitable for LLM fine-tuning. Currently, MedEduChat accepts only text-based inquiries and does not support multimodal inputs such as images, documents, or audio. Establishing clear borders for what LLMs can and cannot do is essential for their effective adaptation to the prostate cancer domain.

Our study sample consisted primarily of older, White, and highly educated prostate cancer patients from Mayo Clinic, which limits the generalizability of the findings to more diverse populations. Three participants declined to share their chat histories, resulting in partial data. Therefore, as the number of unshared chat histories was small, sensitivity analyses excluding these missing cases were not feasible. Recruitment challenges, including participants' age, variable familiarity with conversational AI, the requirement for Mayo Clinic portal access, and the high-stress clinical environment, further constrained our sample diversity. Future work should evaluate MedEduChat with diverse racial or ethnic groups, education levels, and geographic backgrounds to ensure the tool generalizes to all patient groups.

Since the current study was conducted as a single-arm feasibility assessment, the absence of a comparator, such as standard patient education materials or an alternative LLM agent, limits our ability to isolate the effects of the chatbot from potential novelty or observer influences. A future randomized controlled trial [40], incorporating a larger and more heterogeneous sample, will enable direct comparison with conventional educational tools and support a more rigorous evaluation of the chatbot's impact on knowledge retention, user engagement, and health communication outcomes. Moreover, as the study required a researcher moderator to oversee sessions, participants may not have experienced the MedEduChat agent in a fully naturalistic or unobserved setting, introducing the possibility of observer bias. Future study designs should aim to mitigate this bias by enabling more private, remote, or self-directed interaction where appropriate.

Additionally, our study was specifically focused on the use of an LLM-based educational chatbot by non-metastatic prostate cancer patients, and the results may not be generalizable to other types of illness. Some of the topics and concerns that emerged during our research may potentially be useful for broader healthcare technology development, but caution should be used in overgeneralizing the results (Table 2). Future researchers should consider the unique characteristics of different illnesses before extending our approach to other medical contexts.

**Table 2.** Potential Risks in MedEduChat Agent Development & Evaluation

| Risk Category | Explanations |
|---|---|
| **Decision Ambiguity** | **Response Uncertainty:** The LLM agent may generate answers that were hedged or internally inconsistent, particularly when addressing uncommon treatment side-effects or atypical laboratory findings, resulting in guidance that lacked clarity or actionable direction. |
| | **Decision Authority:** When multiple evidence-based options exist, the LLM agent may lack a clear rationale for its recommendation, leading patients to perceive the LLM guidance as arbitrary rather than personalized. |
| | **Remote or Unmonitored Off-Scope Questions:** Without real-time oversight, off-scope patient questions may receive generic disclaimers instead of being properly triaged to appropriate clinical services. |
| **Personalized Response** | **Personalization Limitations:** LLM chatbot can be over-generalized based on demographic features embedded in the training data, offering education that fits population averages rather than individual circumstances. |
| **Underlying Unreliable EHR Data** | **Incomplete or delayed patient EHR data** could directly cause the LLM chatbot to communicate wrong clinical information. |
| | **Incorrect EHR data,** such as staging, lab results (i.e., PSA values), or medication lists, can generate conflicting dosing reminders or follow-up schedules. |

In this paper, we presented a mixed-method usability study with an LLM agent, MedEduchat, for prostate cancer patient education. The study provides a proof of concept demonstrating how such an agent can be integrated into clinical workflows to address patients' questions in a timely and understandable manner. While our findings suggest that MedEduChat can facilitate patient engagement with prostate cancer–related information at their own pace, we interpret these results within the scope of usability evaluation rather than clinical outcomes. Our study contributes to the growing field of agentic healthcare AI by demonstrating how conversational agents can support patient learning in real-world clinical settings. MedEduChat can move beyond static educational materials and offer personalized and interactive support by showing how an LLM agent can be integrated into existing clinical workflows to address patients' questions in a timely, relevant, and understandable way. By supporting patients in navigating complex medical information, MedEduChat reflects a broader shift toward more responsive, patient-centered technologies in healthcare.

## Methods

The study was classified as a quality improvement project, exempt from Mayo Clinic IRB approval (ID: 24-013856). The study followed the Consolidated criteria for reporting qualitative research (COREQ) reporting guidelines [41].

### 1. MedEduChat Agent Design

The educational model that we adopted is known as "5E," including the steps of Engage, Explore, Explain, Elaborate, and Evaluate [42–44]. To apply this model in the prostate cancer education context, we synthesized the five steps into three main stages: health outcome explanation, learning enhancement, and engagement. The health outcome explanation component offers lay-language information to help users understand their diagnosis and treatment options. Users can further explore the details of treatments through the learning enhancement options, depending on their level of interest and literacy. Ultimately, the goal is to help patients engage in their treatment and promote agency and understanding (Figure 3).

MedEduChat incorporated details of a particular user's case, including the cancer type, stage, lab results, treatment history, symptoms, medication history, family medical background, and demographic data. If the patient has already received treatment, then MedEduChat also incorporates relevant treatment details. MedEduChat followed specific instructions defined in its system prompt on how to retrieve data for a patient, given the patient ID. Four data retrieval functions (keyword-based, not Retrieval-Augmented Generation (RAG)/vector search) were defined for this work: get_patient_details(), get_patient_treatment_details(), get_patient_diagnosis_details(), get_patient_clinical_notes(), and get_patient_treatment_details(). When MedEduChat initiated the call to process these functions, based on the context of the conversation with the patient, a data server uses SQL to query the specific information that was requested by the patient. The returned data was then added to the conversation history so that MedEduChat can use it as context. In instances where information was missing, incomplete, or delayed due to EHR update lags, the system omitted these data points. Designed as a long-term conversational agent, MedEduChat progressively integrates updated information over time to ensure continuity and comprehensiveness of patient education.

The retrieval functions rely on keyword-based indexing, using deterministic rules suited to structured patient data. Unlike general retrieval-augmented generation [45], which extracts text from large unstructured corpora, this task-specific approach is more efficient and avoids unnecessary complexity. As LLMs expand context capacity and reduce inference costs, the need for general-purpose RAG diminishes further. MedEduChat data retrieval was designed with the recognition that although patient data is often unstructured, it is still well-organized and indexed. Clinical notes, for example, are dated, signed by a provider, and include a variety of labels that allow for identifying them from a database. If a prostate cancer patient is being treated by radiation oncology, then the LLM ought to be provided with all the radiation oncology and urology clinical notes associated with that patient.

To manage long conversations, MedEduChat implemented defined retrieval functions and a structured pruning procedure. The system prompt remained fixed, while older conversation segments were truncated in fixed-size batches until the total token count was within the predefined limit. Because retrieval functions returned data in reverse chronological order, pruning began with the oldest content. Messages below a specified threshold were preserved, and thresholds were progressively adjusted until the desired balance was achieved. Parameters were configured as follows: maximum conversation length = 95,000 tokens, pruning batch = 250 words, minimum threshold = 10,000 tokens, and decrement = 2,500 tokens. This process prioritized the retention of recent and relevant information while maintaining computational efficiency.

After compiling an overview of the patient's health conditions, MedEduChat offered pre-set prompts to guide the patient toward asking questions that are most relevant to their interests. (These semi-structured pre-set prompts are listed in Supplementary Table 1).

The prompts for MedEduChat were developed through an iterative design process to optimize outcome quality, safety, user-centered, and contextual accuracy. The final prompt framework was organized into four sections: “Purpose,” “Functions,” “Instructions,” and “Resources.” The “Purpose” section established the context for the MedEduChat interaction, while the functions section facilitated the retrieval of EHR-integrated data. Then, the “Functions” section defined nine functions, including patient demographics, diagnoses, treatments, clinical notes, radiology reports, pathology reports, in-basket messages, and appointment details. The “Instructions” section ensured that MedEduChat consistently served an educational role. In cases where users posed unrelated questions, the prompts encouraged exploratory dialogue but redirected the interaction toward the primary educational goals, thereby maintaining both depth and breadth of prostate cancer education coverage. The final “Resources” section specified relevant libraries and webpages that enabled MedEduChat to retrieve accurate and evidence-based information.

MedEduChat prompts’ integrated both closed-domain information, capturing essential patient details, and semi-structured guidance informed by a sustainable education model. The agent emphasized three core goals: providing detailed explanations of health outcomes, enhancing patient learning, and promoting engagement by soliciting user feedback on the quality of interactions. The prompts were validated through multiple rounds of internal testing and refinement to validate their effectiveness in safety, accuracy, and an empathetic conversational tone.

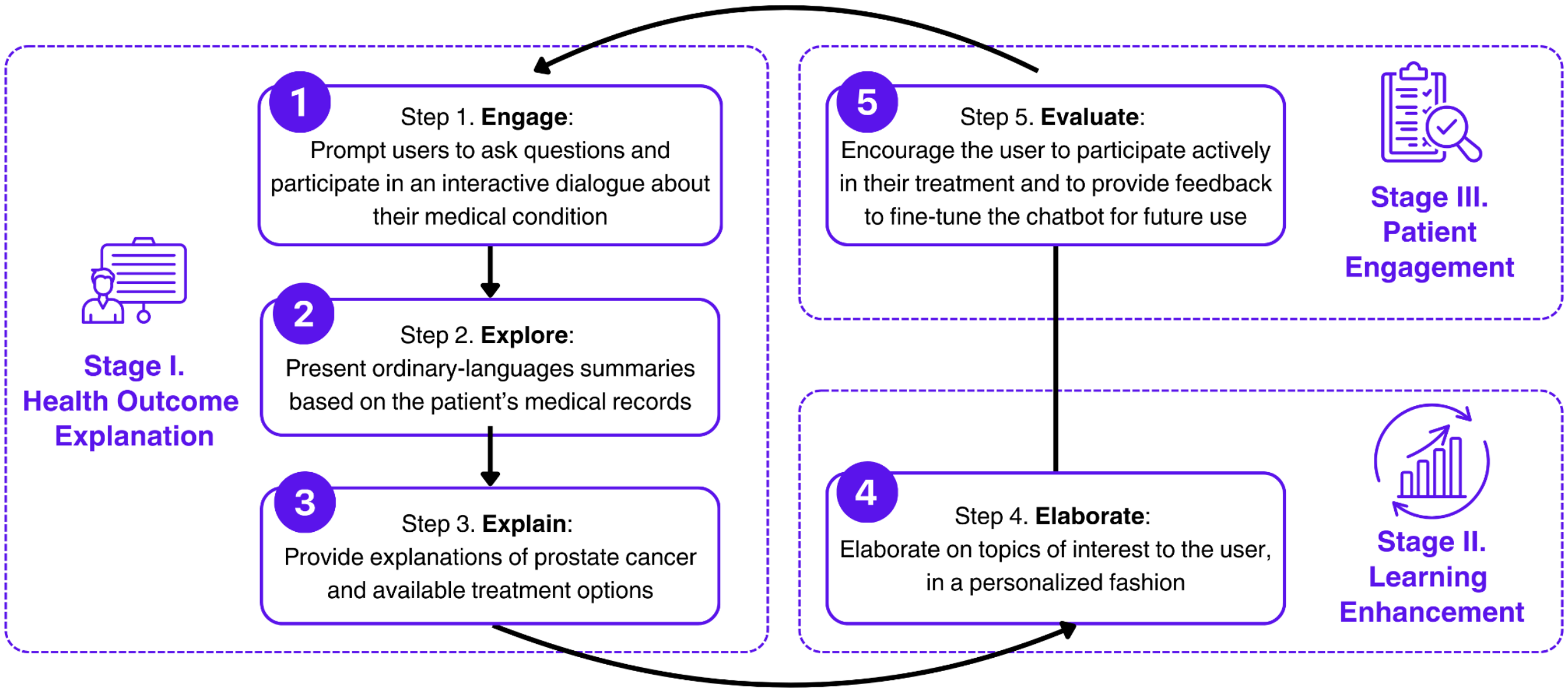

**Figure 3. Stages and steps of the LLM-based chatbot design.** The process consists of three stages: Health Outcome Explanation, Learning Enhancement, and Patient Engagement, with five sequential steps that guide user interaction, personalize information delivery, and incorporate feedback for continuous improvement.

Through connecting with the patient’s EHR profile, MedEduChat was able to incorporate the retrieved patient data as part of the educational session. Upon reviewing the answers, patients can choose to explore more in-depth details or modify their prompts if desired. The goal of these step-by-step interactions was to help patients gain a clearer understanding of their cancer diagnosis, treatment options, treatment history (if any), and potential side effects (Figure 3, Stage I).

During the "Learning Enhancement" stage, patients can learn about their treatment experience or weigh trade-offs in treatment burdens and statistical health outcomes (Figure 3, Stage II). For example, if a shorter distance to the treatment location is a priority for the patient, then they can assess this factor in relation to treatment efficacy. This iterative assessment process enhanced the learning experience and improved patients' comprehension of clinical decisions.

The final "Engagement" stage encouraged patient users to continue taking the initiative in their treatment process and provided an opportunity to evaluate MedEduChat's performance, including any suggestions to help developers fine-tune the chatbot (Figure 3, Stage III). Users also had the option to save or print their conversation history for future reference, and they can receive LLM-generated summaries that highlight key learning points and outcomes. Patients could return and continue interacting with MedEduChat for more information and ask questions as their treatment journey continues. All patients' anonymized chat histories are available in the Supplementary Information.

## 2. MedEduChat Intervention Details

MedEduChat integrates with OpenAI's GPT-4o through a HIPAA-compliant Azure-hosted endpoint provided by Mayo Clinic, ensuring patient privacy and data security while enabling access to necessary health information. Using retrieval functions, MedEduChat retrieves data from the Mayo Clinic's backend EHR systems, including Epic (Mayo Clinic-wide EMR system) and Aria ver. 16 (the oncology information system for radiation oncology). We incorporated information such as family and medical history, lab results, diagnosis details, treatment details, and clinical notes. MedEduChat does not interpret imaging or radiology reports, as it assumes that staging and related findings have already been determined by the clinical team. From the EHR system, we also retrieved patient self-reported education details and patient portal in-basket messages.

MedEduChat is a closed-domain, LLM agent designed to deliver accurate, guideline-aligned, and personalized prostate cancer education using patients' existing health data (Figure 4). By analyzing individual medical records, it generates patient-centered learning materials tailored to literacy, tone, and detail preferences. Unlike public search engines that may expose users to unverified or misleading information, MedEduChat is grounded in institutionally approved educational content and National Comprehensive Cancer Network (NCCN) guidelines, thereby maintaining clinical relevance and minimizing the risk of hallucinations. Its semi-structured design encourages patients to engage in full educational cycles, balancing guidance with flexibility. This approach ensures that users receive accurate, personalized information while preserving the clinician's role in care. MedEduChat adapts dynamically to users' needs throughout each educational session. The MedEduChat technical details and prompts are presented in Supplementary Information.

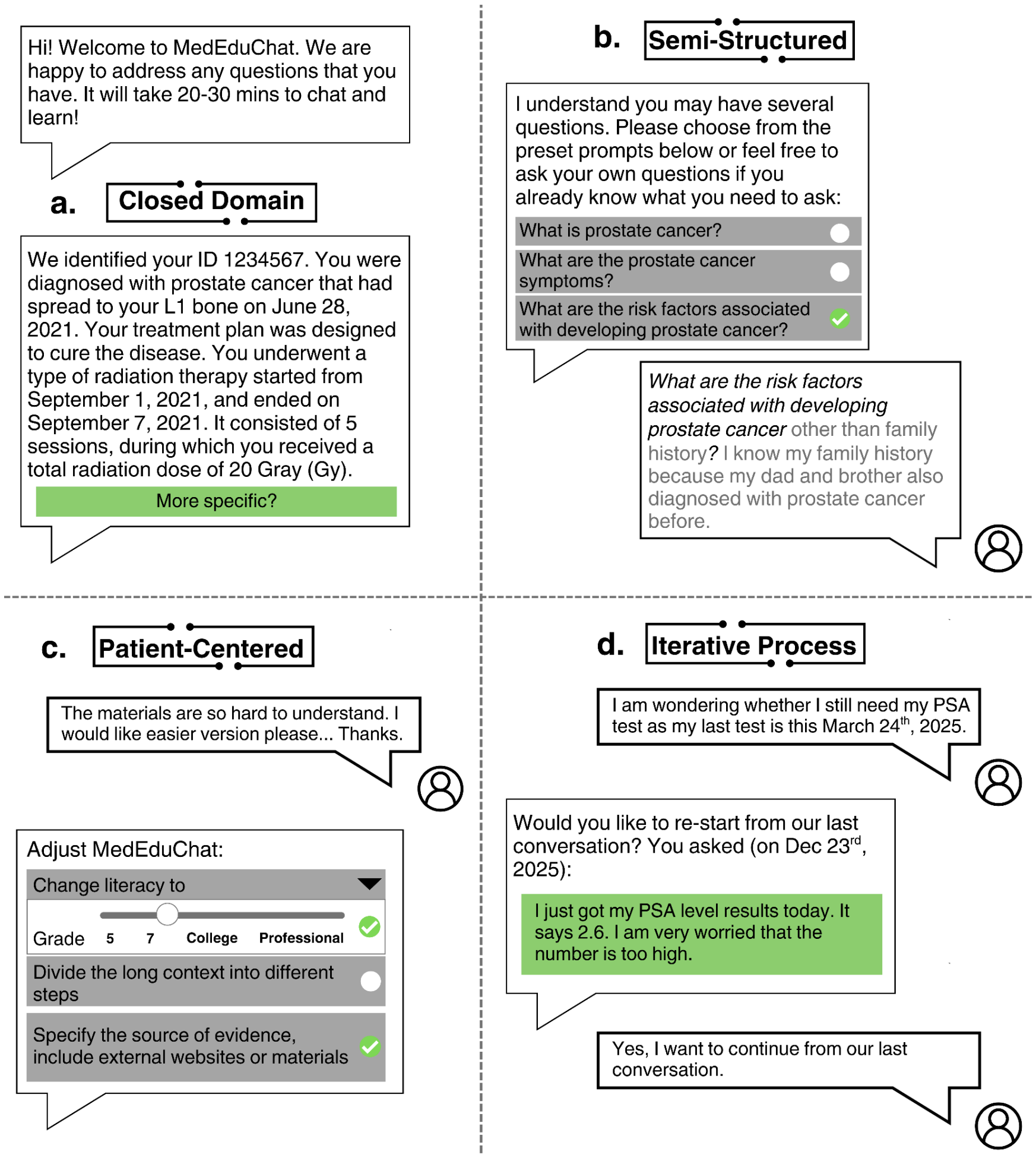

**Figure 4. Features incorporated in the LLM-based MedEduChat design.** The agent supports multiple interaction modes, including closed-domain access for tailored clinical information, semi-structured guidance for question selection, patient-centered personalization, and an iterative process to promote sustained education. These modes enable users to navigate, customize, and engage with the MedEduChat agent according to their needs and preferences.

Following specific system prompt instructions, MedEduChat accesses patient data based on the Mayo Clinic's patient ID. MedEduChat relies on a rule-based keyword retrieval mechanism, rather than employing vector-based retrieval methods such as BM25 or approximate nearest neighbor (ANN). This approach prioritizes clinical interpretability and precision to ensure the retrieved content is explicitly linked to known medical terms and data fields, which reduces the likelihood of semantically related but clinically irrelevant outputs. We only retrieve patient data, process it into LLM-readable text, and provide it in the conversation history. Some retrieval functions rely on keyword matching. When a patient requests specific information, a data server uses SQL to query the relevant records, adding the retrieved data to the conversation history so MedEduChat can incorporate it seamlessly into patient interactions. The full interaction is illustrated in the Supplementary Movie 1.

### 3. Usability Study's Setting

To evaluate the usability, feasibility, and acceptance of MedEduChat, we conducted a mixed-method study with prostate cancer patients recruited from the Mayo Clinic Radiation Oncology departments at two campuses: Phoenix, AZ, and Rochester, MN. A power analysis was conducted to evaluate the statistical sensitivity of our sample size. With 15 participants, a paired-sample design, a two-tailed alpha of 0.05, and 80% power, the study is powered to detect an effect size of approximately Cohen's d = 0.75. While this sample size limits the detection of small to moderate effects, it is sufficient to identify changes in patients' education experience, providing evidence regarding the impact and utility of the MedEduChat agent.

Researcher Y.H. conducted all the studies with the fifteen prostate cancer patients (mean age 74.6 yrs; SD 7.6) (Table 3). The inclusion criteria were that participants had a formal diagnosis of prostate cancer (any stage), that their cancer was non-metastatic, that they had a basic level of computer literacy, and that they were willing to engage with an experimental educational chatbot agent. Participants were asked to sign a consent form before the study, which explained that they would be asked to share their chat history and survey results. The sessions were conducted individually for each participant and took approximately 45-60 minutes to complete. Participants who completed their session received a $30 compensation for their time and contributions. Figure 5 shows an overview of the usability study design.

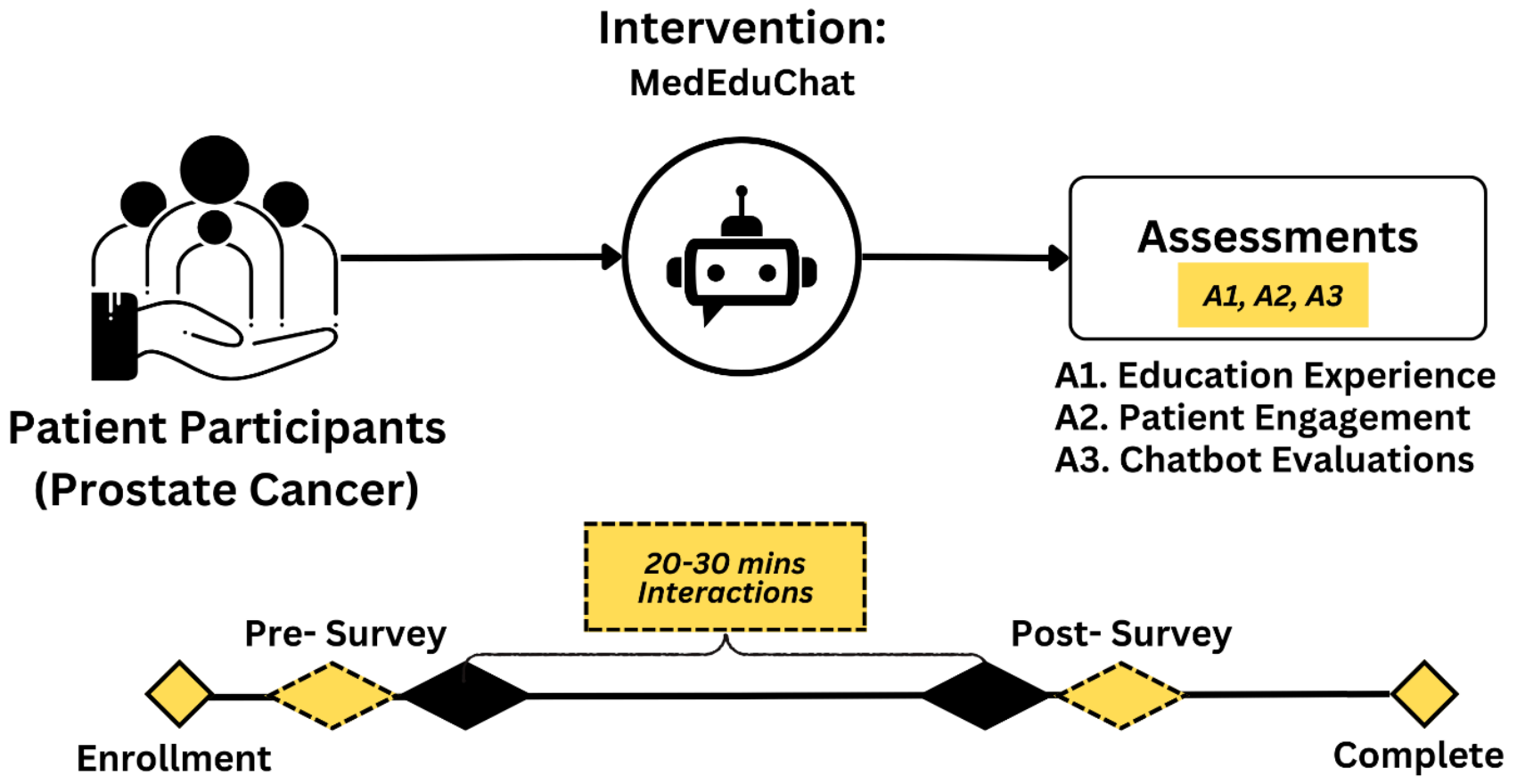

**Figure 5. Usability study design.** Prostate cancer patients interacted with MedEduChat for 20 to 30 minutes, with pre- and post-surveys used to assess education experience, engagement, and chatbot performance.

**Table 3.** Demographic profiles of patient participants in the MedEduChat's usability study.

| Patient Participant ID | Age | Education Level | Race/Ethnicity | Prostate Cancer Stage |
|---|---|---|---|---|
| P1 | 84 | Ph.D. | White | Stage IIB |
| P2 | 80 | M.S. | White | Stage III |
| P3 | 67 | M.S. | White | Stage II |
| P4 | 68 | Ph.D. | White | Stage III |
| P5 | 71 | M.S. | Asian | Stage II |
| P6 | 76 | Ph.D. | White | Stage II |
| P7 | 68 | D.D.S. | White | Stage IV |
| P8 | 81 | B.S. | White | Stage IIC |
| P9 | 85 | B.S. | White | Stage IIIB |
| P10 | 72 | M.S. | White | Stage IIB |
| P11 | 68 | M.S. | White | Stage IIIB |
| P12 | 63 | B.S. | White | Stage IIB |
| P13 | 46 | M.S. | White | Stage IVA |
| P14 | 70 | B.S. | White | Stage IIB |
| P15 | 73 | B.S. | White | Stage IIIB |

The participants were first asked to fill out a pre-intervention survey, which took approximately 6 to 10 minutes. They were then asked to engage in interactions with MedEduChat for approximately 20 to 30 minutes. Finally, they were asked to complete a post-intervention survey and provide feedback about the agent, which took 15 to 20 minutes. Since MedEduChat is still a developing prototype and may generate errors, users accessed it through screens shared by the researchers, allowing us to monitor the interactions and intervene if necessary. However, the patients were allowed to interact freely with MedEduChat without external instructions, and interruptions occurred only if a usability problem, misinformation, or bug was encountered. We acknowledge that screen-sharing may have introduced a potential Hawthorne effect [46], as participants were aware that their interactions were being observed. All interruptions and researcher interventions were recorded, including their timing and reason.

The pre- and post-intervention surveys were structured to capture baseline characteristics and subsequent changes following system use. The pre-intervention survey included demographic information, estimated time spent on prostate cancer education before the study, a 4-item health confidence score scale (HCS) [47,48], and a 5-item patient health engagement (PHE) scale [49]. The post-intervention survey repeated the HCS scale to capture any changes in confidence, presented the new 4-item Usability Metric for User Experience (UMUX) [50], and 5 questions regarding participants' interaction experience with MedEduChat. We administered the PHE scale once in the pre-intervention survey to characterize each patient participant's pre-existing engagement with their health. Our intervention consisted of a single, time-limited interaction with the MedEduChat agent, which was not intended to modify trait-level engagement. To limit respondent burden and avoid demand characteristics from repeating the same reflective items immediately after a brief exposure, we treated baseline PHE as a contextual covariate rather than an outcome. Additionally, UMUX was not included in the pre-intervention survey, as it is specifically designed to evaluate perceived usability after system use. Participants were also welcome to provide some open-ended comments about their favorite and least favorite parts of the interaction with MedEduChat. This design allowed us to compare baseline engagement and confidence with post-intervention outcomes while also assessing perceived usability and user experience. Full survey instruments are provided in the Supplementary Information.

The three main qualitative themes were **patient education experience**, **patient engagement**, and **agent chatbot efficacy**. Three clinical professionals independently evaluated the chat interactions (Table 4) across six dimensions, each rated on a 1–3 scale (1 = lowest, 3 = highest). (1) "*Answer Correctness*" examines whether the LLM-generated response is factually and clinically accurate based on evidence-based guidelines. (2) "*Patient Question Difficulty*" assesses the complexity of the patient's inquiry, considering both medical depth and communication challenges. (3) "*Answer Completeness*" measures how thoroughly the response addresses all relevant aspects of the patient's question with sufficient contextual detail. (4) "*Patient-Readiness*" evaluates the clarity, empathy, and accessibility of the response to ensure it is understandable and appropriate for direct patient communication. (5) "*Safety and Harm Potential*" considers whether the content poses any risk of misinformation, omission, or advice that could negatively impact patient safety. (6) "*Patient-Centeredness Response*" assesses how well the response is tailored to the specific patient's context and concerns, rather than providing overly general or templated information. To account for rater disagreement, we calculated the mean score across the three raters for each dimension. This approach reflects the general agreement across graders while accounting for minor differences in judgment.

The qualitative commentary provided by these experts was audio-recorded, transcribed, and summarized, and each clinician received $70 in compensation for their assistance. The clinical professionals' detailed grading instructions can be found in the Supplementary Information.

**Table 4.** Demographic profiles of clinical professionals who were asked to evaluate the MedEduChat interaction data during the usability study.

| Clinical Grader ID | Clinical Domain | Years of Experience |
|---|---|---|
| C1 | Medical Physics | 18 yrs |
| C2 | Radiation Oncology | 4 yrs |
| C3 | Prostate Cancer Education (i.e., anxiety disorder, depression) | 12 yrs |

**4. Data Analysis**
Interviewed transcripts were anonymized and analyzed in Google Docs and Google Sheets for thematic analysis and axial coding [51]. Five authors (Y.H., J.M.H., M.R.W., K.V., and W.L.) identified the initial themes, with Y.H. and J.M.H. conducting line-by-line coding to inductively generate initial concepts. These concepts were then grouped into overarching themes and subthemes, with conceptual links identified among them. A consensus on the final themes was reached in collaboration with N.Y., H.P., D.M., C.E.L., A.V., S.K., M.G., and W.L.

All quantitative analyses were conducted using descriptive and inferential statistics to evaluate patient usability outcomes and clinician evaluations of MedEduChat. Descriptive statistics included means, SD, and 95% CI for all continuous measures. For the Patient Usability Study, the PHE score and HCS were summarized to describe participants' baseline engagement and confidence levels. Changes in HCS before and after using MedEduChat were assessed using a Wilcoxon signed-rank test to account for the small sample size and non-normality of paired data.

The analysis of clinical professionals' evaluations of participants' chat histories was conducted by Y.H. The scores of three clinical raters were aggregated across six evaluation dimensions. For each dimension, means, SDs, and 95% CIs were reported. Inter-rater variability was accounted for through averaging procedures, and descriptive summaries were visualized using boxplots.

## Data Availability

We released 85 pairs of anonymized patients' chat histories with MedEduChat in Supplementary Data 1. Also, a sample chat history and a sample interaction video are provided in the Supplementary Information. We preregistered our data collection plan for a future randomized clinical trial on the Open Science Framework under the title "EHR-Integrated Large Language Model Agent for Personalized Prostate Cancer Education: A Randomized Controlled Trial" (available at https://osf.io/q8j36).

Codes used for data processing and analysis were provided in the Supplementary Information. The full set of prompts, including retrieval functions, is also available in our public GitHub repository at: https://github.com/YuexingHao/MedEduChat.

## Acknowledgement

This research was supported by the American Psychological Foundation K. Anders Ericsson Dissertation Research Grant, President's Council of Cornell Women (PCCW) Frank H.T. Rhodes Leadership Grant and Mission Grant, Alan D. Mathios Research & Service Grant, National Cancer Institute (NCI) R01CA280134, the Eric & Wendy Schmidt Fund for AI Research and Innovation, the Fred C. and Katherine B. Anderson Foundation, and the Kemper Marley Foundation. The funders had no role in study design, data collection and analysis, decision to publish, or preparation of the manuscript.

## Competing Interests

The authors declare that they have no competing interests.

## Author Contributions

Ms Hao, Dr Holmes, Dr. Waddle, and Dr Liu had full access to all of the data in the study and take responsibility for the integrity of the data and the accuracy of the data analysis.
**Concept and design:** Hao, Holmes, Waddle, Liu.
**Acquisition of data (including patients' recruitment and providing patients' clinical information):** Hao, Holmes, Waddle, Davis, Yu, Liu.
**Analysis and interpretation of data:** Hao, Holmes, Liu.
**Drafting of the manuscript:** Hao, Holmes, Liu.
**Critical review of the manuscript for important intellectual content:** Hao, Holmes, Waddle, Davis, Yu, Vickers, Preston, Margolin, Löckenhoff, Vashistha, Kalantari, Ghassemi, Liu.
**Statistical analysis:** Hao.
**Obtained funding:** Waddle, Ghassemi, Liu.
**Administrative, technical, or material support:** Hao, Holmes, Waddle, Davis, Yu, Vickers, Preston, Ghassemi, Liu.
**Supervision:** Waddle, Davis, Yu, Margolin, Löckenhoff, Vashistha, Kalantari, Ghassemi, Liu.